# Application of Principal Component Analysis and Artificial Neural Networks for the Prediction of QoS in FSO Links over South Africa

## S.O Adebusola [1], P.A Owolawi[1], J.S Ojo[3], P.S Maswikaneng[2], A.O Ayo[1],

**[1]Department of Computer Systems Engineering, Tshwane University of Technology, Pretoria 0001, South Africa.**

**[2]Department of Information Technology, Tshwane University of Technology, Pretoria 0001, South Africa.**

**[3]Department of Physics, Federal University of Technology, Akure 340110, Nigeria.**

**Abstract-** Optical Communication in Free Space (FSO) bids more radio bandwidth, operates under a gratis license, and has a lower startup cost as compared to Radio Frequency (RF). Nonetheless, its vulnerability to variations in atmospheric meteorological circumstances is a concern. Ultimately, the purpose of this study is to use Principal Component Analysis (PCA) with Artificial Neural Networks (ANN) to design a QoS prediction model for an FSO communication connection. To accomplish the specified goal, meteorological data such as visibility, wind speed, and altitude were collected from the Weather Services in South Africa (SAWS) archive during a ten-year duration (2010–2019) at five different locations: George, Johannesburg, Kimberly, Bloemfontein, and Polokwane. The eigenvalues of the first Principal Component (PC1) and the second Principal Component (PC2) in the PCA across the stations Bloemfontein, Johannesburg, Kimberly, George, and Polokwane are 7.624 and 1.020, 7.234, and 0.984, 6.204 and 1.723, 7.354 and 0.876, and 7.104 and 0.865, respectively, demonstrating that, they are kept as QoS variables to train the Artificial Neural Network (ANN) model as they provide the most compelling interpretation of the original variable data. The RMSE values of every proposed model across all the study locations are 0.1437, 0.2131, 0.2329, 0.1101, and 0.1977 respectively. Based on the RMSE, the proposed performed better over George. A realistic and accurate predictive model is developed for each of the study locations. Thus, the developed model will serve as a valuable tool for maintaining good QoS in FSO network services and improving telecom businesses in South Africa.

## 1 Introduction

The term Optical Communication in Free Space (FSO) describes every method based on achieving high-bandwidth communication via the environment by transmitting controlled optical signals. Fiber-optical transmission technology and FSO operate on precisely the same theoretical foundation. The alteration is that instead of being guided by an optical cable, the intensity beam is nearly parallelized and directed from the origin to the end location via the atmosphere (Mustafa *et al*., 2011; Adebusola *et al*., 2021; Maswikaneng *et al*., 2023). The telecommunications sector has taken notice of FSO because of its affordability, ease of installation, swift connection establishment, particularly in emergency scenarios, high broadband supplies, and variety of uses. In contrast to Radio Frequency (RF) technologies for communication, which have a permissible transmitted information speed of 622 Mbps, FSO transmission allows for maximal transmission speeds as high as 2.5 Gbps. (Alkholidi and Altowij, 2012; Sharoar *et al*., 2020). Authors (Maswikaneng *et al*., 2018; Ojo *et al*., 2018; Maswikaneng *et al*., (2022), and Katiyar (2013) found that vapor absorption, precipitation dissipate cloud, foggy diminution, and tropospheric vibrating were the main air and weather-related causes of transmission loss. Fog attenuation is often linked to signal absorption and dispersion throughout the channel, particularly at the receiver's end. Aside from the impact of meteorological conditions on the availability of the FSO connection, further issues include environmental light source disruption, laser contrast, topological reduction, block and shadowing, placement, and monitoring. (Maswikaneng *et al.*, 2023). The predominant climate conditions vary depending on the place. Fog and rain attenuation result in complete signal loss at the receiving terminal. This is a result of the receiving terminal's inability to adapt to every circumstance by itself. Additionally, FSO replication stations in the receiver station were not designed to deal with such situations. (Ojo *et al.*, 2018). Accordingly, it is important to plan a system that will logically maintain the condition once the need arises. Based on the statistical analysis of the weather visibility spread as well as various attenuations, the calculation of Quality of Service (QoS) at various locations and distances has been analyzed. The system designer becomes interested in the Quality of Service (QoS) measure to take into consideration the signal reliability and accessibility of the system. The main issue facing FSO services is how weather conditions affect the communication channel. The transmission connection and weather circumstances might lead to a decrease in the Signal-to-Noise ratio (SNR), which can provide additional issues. For FSO systems, the distance of transmitting signals from the point of origin to the endpoint is extensive. (Maswikaneng *et al.*, 2023). For many FSO operations and providers of services, it is therefore extremely vulnerable to climatic reduction induced mostly by weather conditions. (Maswikaneng *et al.*, 2018; Ojo *et al.*, 2018).

The QoS in the FSO serves a significant role in ensuring client happiness as well as an evaluation of system functionality regarding the Client Service Level Agreement (CSLA) of the service provider and the clients they serve (Lu, 2002). Reliability and long-term viability of the SLA by QoS might be achieved by a system's capacity to intelligently react to data rate, modulation, signal strength standards, and coding. The measurements concerning network efficiency had been computed under various circumstances. The authors (Ojo *et al.*, 2018) have also mentioned that there is a lack of widespread availability for approaches that aim to integrate this prediction model in a linked way. Consequently, it is imperative to accurately identify and predict the overall impact of all significant fog extenuating factors on quality of service (QoS), including site, transmitted, and dissemination features within any specific route among FSO transmitting devices. Empirical approaches are carried out by evaluating the combined effects of several attenuations despite having a thorough understanding of their occurrence probability. Furthermore, whenever several QoS variables show multicollinearity or when the variable information is related to each other, it adversely affects the precision of predictions in a model. This makes using regression models more difficult in many ways. Eliminating one or more variables that have a larger autocorrelation amongst the variables and are less explanatory of the dependent variable is the easiest way to solve this problem (Chan *et al.*, 2015). Moreover, by using the variables that remain to create a prediction model, multi-collinearity impacts are efficiently avoided, and the model's forecasting reliability is increased. Meanwhile, to lower the total number of variables, researchers are now using stepwise selection (SR), forward selection, optimization subclass selection, and backward selection. The SR methods, as compared to other methods indicated above, are presumably among the most applied study techniques in studies of substance and efficacy (Chan *et al.*, 2015). One benefit is that it saves time. In their research, (Chan *et al*., 2015) concluded that the SR technique had flaws, including biased results and process bias. This research study aims to investigate an accurate and practical variable sorting technique in light of the shortcomings of the SR technique to address the multicollinearity issue of the information while retaining as much of the variable information as possible and enhancing the accuracy prediction model. One such analysis technique that can successfully reduce parameter dimensions while retaining effective influence components in QoS parameters is the analysis of principal components (PCA). The PCA approach is still usually employed today for face recognition (Kanwal and Shahid, 2013), image compression (Rujirakul *et al*., 2016), and image representation (Zhu *et al*., 2015) (Arafah and Moghli, 2017). Nevertheless, it hasn't gotten much use or attention when it comes to resolving multicollinearity issues in data and raising model prediction accuracy. Moreover, the analysis of the consequence of visibility on FSO-to-Satellite link availability was studied over some selected locations: Ikeja, Akure, Jos, Enugu, Sokoto, and Port-Harcourt (PH), Nigeria, and the results showed that at 352 THz frequency, Ikeja, PH, and Jos had a link margin of -109.77, -91.99, and -37.48 dB, respectively (Falodun *et al*., 2020). Also, authors (Ojo and Owolawi 2015) developed an intelligent system to tolerate the QoS at Super High Frequency (SHF) and Extreme High Frequency (EHF) satellite system networks in a subtropical climate. Furthermore, authors Chavan and Patane (2015) used a fuzzy logic decision system to create an enhanced intelligent climate scheme for cable networks. The results obtained show that the QoS of satellites could be improved with fuzzy logic in the system. In addition, Chen *et al*. (2020) created a model to forecast the cost of an overall aeronautics airplane using the ANN and PCA. The model is tested and trained using 22 overall aeronautics airplane samples in total. The MAPE, MAE, R, and RMSE were adopted as the performance metrics for every training along with testing samples. The values of the metrics MAPE, MAE, R, and RMSE obtained for every training and testing sample are 0.0009 and 0.015, 1.222 and 3, 0.9999 and 0.9994, and 1.667 and 3.146, respectively. To our understanding, there are no current works that have described the amalgamation of ANN and PCA approaches to resolve the glitches of QoS in free space optic communication systems. Hence, a hybrid ANN-PCA-founded technique can be measured to forecast the reliability of QoS in FSO communication links. The purpose of this study is to forecast the reliability of QoS in FSO links over a subtropical climate using a hybrid PCA-ANN model. Also, verify the influence of the atmospheric diminution factor on the reliability of QoS. The remaining sections of this work are arranged as follows: Section II covers the technique; Section III covers the results and discussion; and Section IV covers the conclusion and future study.

II                                    Methodology

2.1 Data collection

The meteorological visibility information was downloaded from the annal of the Weather Services of South Africa (SAWS) over ten years (January 2010–December 2019) over Bloemfontein, Johannesburg, Kimberley, George, and Polokwane, South Africa. The information was downloaded three times a day (8h00, 14h00, and 20h00.). The atmospheric attenuation coefficient was assessed by taking the average of the data collected over ten years. Using the outcome, the ratio of signal to noise (SNR) for different optical wavelengths (760, 860, 960, 1260, and 1550 nm) was calculated. The climatic variation of the study areas is explained in Kolawole *et al*. (2022). Figure 1 shows the location where the data was collected, and the operational settings used to simulate the results are displayed in Table 1.

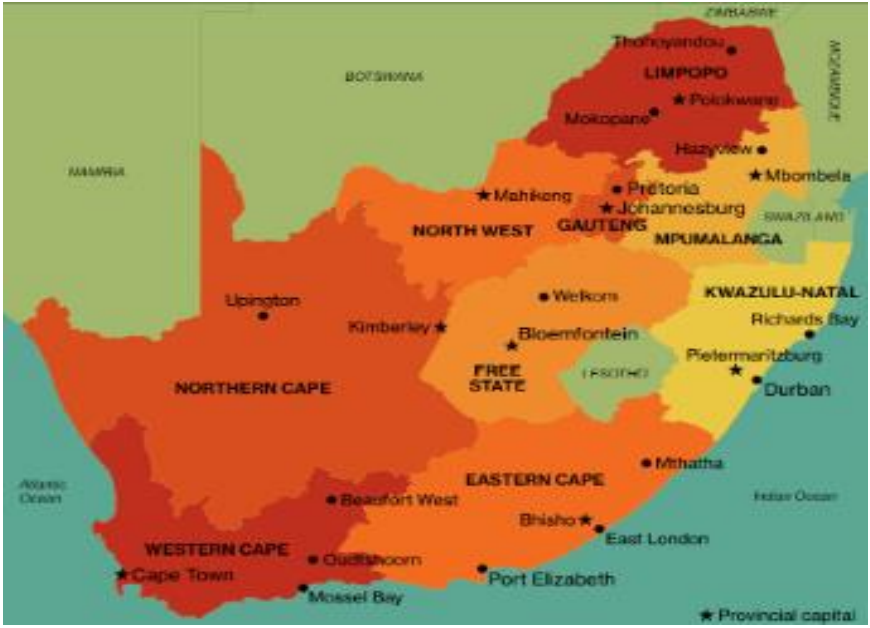

Figure1 shows the provinces where the data was collected (Kolawole *et al.*, 2022)

Table 1: Operation parameters of an FSO System

| Operating parameters | Values |
| --- | --- |
| Transmitter power | 5-100 mW |
| Divergence Angle | 3 mrad |
| Efficacy of the transmitting antenna | 80% |
| Efficacy of the receiving antenna | 80% |
| wavelengths | 760,860,960,1260, and 1550nm |
| Range | 0.1≤L≤1 km |
| Visibility | location dependent |
| Receiver Sensitivity | -40 dBm |
| Data rate | $1 \times 10^9$GB |
| PIN Load resistance | 1000 ohms |
| Boltzmann constant | 1.36E-23 J/k |
| Absolute Photodiode Temperature | 298 K |
| Dark Current | 10 nA |
| Responsivity | 0.7A/w |
| Electrical Bandwidth | 0.5 GHz |

## 2.2 Modelling of Optical Beam Propagation via the Atmospheric Channel

Beer-Lambert's law describes how optical signals propagate across turbulence-filled air channels. As mentioned in, Maswikaneng *et al.* (2022).

$$\Gamma(\lambda, L) = e^{-(\tau_{ext}, L)} \tag{1}$$

where $\Gamma(\lambda, \text{L})$ signifies the transmittance of the atmospheric network, $\tau_{ext}$ connotes the coefficient of atmospheric attenuation, and L connotes broadcast distance. The transmittance is dependent on the meteorological visibility, $(V)$ which defines the transparency of the atmosphere. The route length at which the transmittance falls to a specific transmission threshold value is known as the visibility (i.e., $\Gamma(\lambda, \text{L}) = T_{th}$). For optical wireless communication systems, the broadcast threshold, likewise referred to as the visual threshold, is set at 2/100, while for runway visual range at airports, it is set at 5% (Kolawole *et al*., 2017). The coefficient of air attenuation, at $T_{th} = 2\%$ is written as follows:

$$\tau_{ext} = -\ln \frac{0.02}{V} = \frac{3.912}{V} \tag{2}$$

whereas V connotes visibility measured in kilometers. The expression in (3) can then be used to compute the attenuation of the atmosphere in dB.

$$\chi(L, V) = 10 \, \log_{10}(e)\tau_{ext}(V) \, L \tag{3}$$

From visible to near-infrared wavelengths, the total extinction coefficient can be calculated by the Kruse formula and is modified (Kolawole *et al*., 2017, Ghassemlooy and Popoola, 2022, and Kim *et al*.,2001):

$$\mu_{\mu,sct}(V) = \beta_{ae}(\lambda) = \frac{10 \log_{10} T_{th}}{V(km)} \times \frac{\lambda(nm)^{-\rho_o(v)}}{\lambda_o} \tag{4}$$

where $\lambda$ is the wavelength measured in nm, $\lambda_o$ is the all-out band wavelength of the solar band, and $\rho_o$ is the magnitude component distribution limit and is written as (Kolawole *et al*., 2017):

$$\rho_o(V) = \begin{cases} 1.6, & if & V > 50 \, km \\ 1.3, & if & 6 \, km < V < 50 \, km \\ 0.585V^{\frac{1}{3}}, & if & 0 \, km < V < 6 \, km \end{cases} \tag{5}$$

Several studies have employed the Kruse model to compute the aerosol scattering factor. Nevertheless, it was found in (Kim *et al*., 2000) that the component magnitude dispersal parameter in the Kruse model cannot effectively estimate the values of the scattering attenuation coefficient for visibility less than 6 km in foggy weather. As a result, (Kim *et al*., 2000) offered a novel change for the particle size-related coefficient to evaluate it further correctly, which is given as follows:

$$\rho_o(V) = \begin{cases} 1.6, & if & V > 50 \text{ km} \\ 1.3, & if & 6 < V < 50 \text{ km} \\ 0.16V + 0.34, & if & 1 < V < 6 \text{ km} \\ V - 0.5, & if & 0.5 < V < 1 \text{ km} \\ 0, & if & V < 0.5 \text{ km} \end{cases} \tag{6}$$

## 2.3        Ratio of Signal to Noise

This is a vital communication connection metric for assessing quality, as it has an inverse relationship to attenuation. (Yasir *et al*., 2018) describe the SNR mathematical expression.

$$Q_{snr} = P_{tmt} - 30 - 10log(G_{tmt}) + 10\log(G_{rvr}) - 20\log(\frac{4\pi}{\lambda}) - 10\log(B_{dth}T_{amb}K_{blt}) - \tau - NR_f - F_{mgn}$$

$$(7)$$

where $P_{tmt}$ means power transmitted, $G_{tmt}$ represents antenna transmitted gain, $G_{rvr}$ the antenna received gain, $\lambda$ wavelength, $K_{blt}$ is the constant of Boltzmann (1.38*10^34 J/K), $B_{dth}$ *is the* received bandwidth (BW = 1 MHz), $T_{amb}$ the surrounding air temperature in K, $\tau$ is the overall diminution factor in dB/km, $NR_f$ represents received figure noise and $F_{mgn}$ represents faded allowance.

**2.4**                                         Artificial Neural Network Application

ANN techniques are complicated net structures that are broadly interrelated by heaps of modest processing components called nerve cells. The most popular ANN techniques in practical applications are those trained using the backpropagation (BP) technique (Chen *et al*., 2019). The gradient steepest descent approach is employed by the BP method to minimize model error while also comparing the goal and output values. Fig. 2 depicts a typical ANN construction.

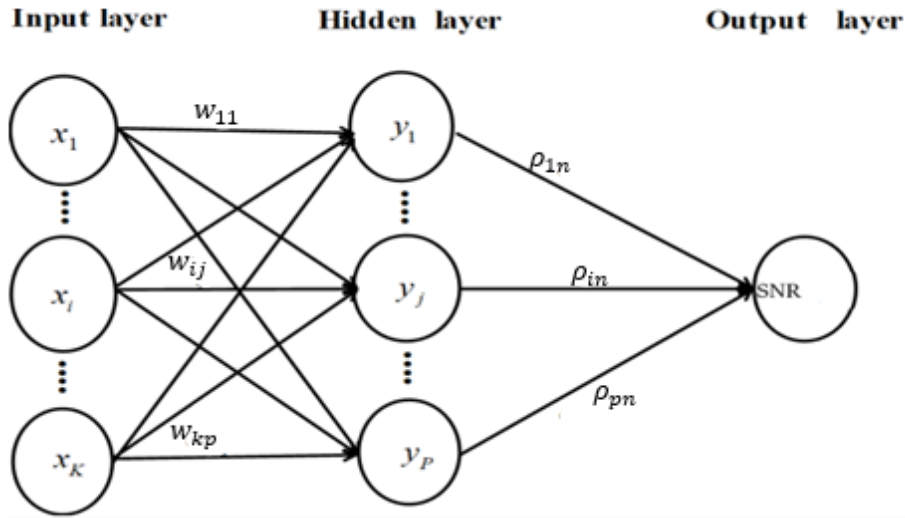

Figure. 2: Artificial Neural Network Layers

The input, hidden, and output layers the three linked layers that make up the ANN technique are seen in Figure 2. Each layer also has a unique initial weight, amount of nerve cells, neural activity, and biases. Neurons in each layer do two things: summarization and activation. To get the output, the weighted input must first be gathered, and then the total must be compressed. Neurons exchange information by way of layers, wherein impulses are received from a single neuron and subsequently sent to the next. To train the ANN model, the BP method is applied. To modify the weight value, the algorithm creates an overall discrepancy between the real and expected results and propagates the error back. Network error reduction is the primary goal of network training. The criterion for ending the training is the mean square error (MSE) ascertained at the output layer (Chen *et al*., 2020). If the mean square error (MSE) rises above a predefined threshold, the ANN model's training is terminated. If not, the concealed and output layers' weights are updated and transmitted backward using the BP approach. Until the MSE value reaches the target value, the weights of the input and concealed layers are adjusted similarly. This achieves the intended outcome of lessening the difference between the expected and desired output values. After training, the testing subset evaluates the network's efficiency. Of the whole data set, 30% was used to assess network performance, fifteen percent was used for validation, and seventy percent was used for network training. (Maswikaneng *et al*., 2023).

## 2.5             Principal Component Analysis (PCA)

A common numerical technique for reducing the number of variables in a data matrix is principal component analysis (PCA), also known as eigenvector-based multivariate analysis. This is a commonly used method for compressing and transforming a training data set's dimension (m-dimension) into a smaller dimension (l). Principal components (PCs), which are uncorrelated variables, are created by PCA after it analyzes a set of related variables in the data set. To do this, data in a p-spatial area is converted into an original dimension d coordinate structure, whereby m ≥ l. (Adisa *et al*., 2019). Reducing dimensionality is typically one of the primary issues in contemporary data analysis. The multidimensional data sets can be challenging to analyze, and it is impossible to immediately understand their framework. Several correlated variables are converted using principal component analysis (PCA) into a reduced number of un-correlated variables known as principal components (PCs), which are sufficient to explain the data structure (Chen *et al*., 2020). PCA's primary goal is to project data into a lower-dimensional space from a higher-dimensional one. Furthermore, PCA facilitates the identification of the specific differences across samples as well as the variables that have the greatest influence on these differences (Alver and Kazan, 2020). The data's variability is explained by the principal components; the first set of PCs is calculated so that they account for the majority of the erraticism in the numbers, after which the rest of the PCs may be disregarded (Hallinan, 2012). The resulting data structure can be immediately and graphically seen by projecting things onto the space that the chosen PCs have defined. PCA facilitates enhanced data visualization and optimizes the learning algorithm's utilization of resources (Lee *et al*., 2015).

## 2.6             The Procedural Minutiae of PCA

Starting with a data collection represented by an n×m matrix, X, where the n columns are the variables and the m columns are the observables, we linearly convert X into another matrix, $\Upsilon$, which is likewise of size $n \times m$, so that for some $n \times n$ matrix, P,

Then,

$$\Upsilon = PX \qquad\qquad\qquad (8)$$

This reckoning denotes a variation of base. Considering the echelons of P to be vectors $P_1, P_2, P_3, \dots P_n$, and the columns of X to be column vectors $x_1, x_2, x_3, \dots x_m$,

$$PX = (Px_1\, Px_2\, Px_3\, \dots\, Px_m) = \begin{bmatrix} P_1x_1 & P_1x_2 & P_1x_3 & \dots & P_1x_m \\ P_2x_1 & P_2x_2 & P_2x_3 & \dots & P_2x_m \\ \vdots & \vdots & \ddots & & \vdots \\ P_nx_1 & P_nx_2 & P_nx_3 & \dots & P_nx_m \end{bmatrix} = \Upsilon \qquad (9)$$

Note, $P_i, x_j \in \mathbb{R}^n$, that and $P_i \cdot x_j$ is the standard Euclidean inner product. PCA establishes independence by analyzing the variance of the original data, determining the maximum variance directions, and then defining the original base accordingly. Reminiscence, the meaning of the discrepancy ($\sigma_Z^2$) of an arbitrary variable (Shlens 2009), $Z$ with average, $\mu$.

$$\sigma_Z^2 = \mathrm{E}[(Z - \mu)^2] \qquad\qquad\qquad (10)$$

Supposing there is a vector of m discrete measu rements, $\tilde{\alpha} = (\tilde{\alpha}_1, \tilde{\alpha}_2, \dots, \tilde{\alpha}_m)$, with $\mu_\alpha$. If the mean of each measurement is being subtracted, then the results obtained could be a translated set of measurements $\alpha =$ $(\alpha_1, \alpha_2, \dots, \alpha_m)$ and has zilch average. Thus, the discrepancy of these dimensions is associated by (Shlens 2009)

$$\sigma_\alpha^2 = \frac{1}{n} \alpha\alpha^t \tag{11}$$

For a second vector of m measurements, $\psi = (\psi_1, \ \psi_2, \cdots, \psi_m)$, which also has a mean of null, then this idea could be generalized to obtain the Covariance of $\alpha$ and $\psi$. Covariance is the degree of how considerably binary variables transformed together. Discrepancy is a different event of co-variance, once the variables are indistinguishable. It's accurate to divide by a feature of $n-1$ instead of n. this is given by                                                                (Shlens                                  2009)

$$\sigma_{\alpha,\psi}^2 = \frac{1}{n-1} \alpha\psi^t \tag{12}$$

We can now generalize this idea to considering our n × m data matrix, X. Remember that n was the number of variables, and m was the number of examples. We can thus ponder on this matrix, X in terms of n row eigenvectors, each of length m.

$$X = \begin{pmatrix} x_{1,1} & x_{1,2} & \dots & x_{1,m} \\ x_{2,1} & x_{2,2} & \dots & x_{2,m} \\ \vdots & \vdots & \ddots & \vdots \\ x_{n,1} & x_{n,2} & & x_{n,m} \end{pmatrix} = \begin{pmatrix} x_1 \\ x_2 \\ \vdots \\ x_n \end{pmatrix} \in \ \mathbb{R}^{n \times m}, x_i' \in \mathbb{R}^m \tag{13}$$

Meanwhile, we have a row eigenvector for individual eigenvalues variables, each of these eigenvectors encompasses all the examples for a precise variable. So, for instance, $x_i$ is an eigenvector of the m examples for the $i^{th}$ eigenvalues. It is then becoming significant to ponder the covariance ($C_X$) as the matrix product given by

$$C_X = \frac{1}{n-1} XX' = \frac{1}{n-1}\begin{pmatrix} x_1 x_1' & x_1 x_2' & \dots & x_1 x_n' \\ x_2 x_1' & x_2 x_2' & \cdots & x_2 x_n' \\ \vdots & \vdots & & \vdots \\ x_n x_1' & x_n x_2' & & x_n x_n' \end{pmatrix} \in \ \mathbb{R}^{n \times n} \tag{14}$$

The PCA method assumes uncorrelated variables in the transformed matrix, $C_Y$, with covariances close to zero. Large variance values indicate interesting system dynamics, requiring $C_Y$, to be constructed with the   requirements listed below:

(i)  Make best use of the signal, measured by variance (maximize the diagonal entries)
(ii). Diminish the co-variance between variables (minimize the off-diagonal entries)

 For the Covariance of $C_Y$ to be estimated, vectors $P_1, P_2, P_3, \dots P_n$ are orthogonal.  Recall from the theorem of linear algebra that the covariance matrix  $C_Y$, can be written in terms of X and P

$$C_Y = \frac{1}{n-1} \ YY' = \frac{1}{n-1}(PX)(PX)' = \frac{1}{n-1}(PX)(X'P') = \frac{1}{n-1}P(XX')P' \tag{15}$$

$$C_Y = \frac{1}{n-1}P\psi P' \tag{16}$$

where  $\psi = XX'$ \tag{17}

$\Psi$  is a $n \times n$ symmetric matrix.

$$(XX')' = (X')'(X)' = XX' \tag{18}$$

Furthermore, remember that every square symmetric matrix is orthogonally diagonalizable is

$$\Psi = EDE' \tag{19}$$

where E, is an n × n orthonormal matrix whose columns are the orthonormal eigenvectors of $\Psi$ and D is a diagonal matrix that has the eigenvalues of $\Psi$ as its diagonal entries.

The rank $\alpha$ of $\Psi$ is the number of orthonormal eigenvectors that it has. By simple transformation of the matrix P, the rows of P are taken as the eigenvectors of $\Psi$, then P = E′ and vice versa, by substitution, the covariance matrix $C_Y$ becomes:

$$C_Y = \frac{1}{n-1}P\psi P' = \frac{1}{n-1}E'(EDE')E \tag{20}$$

Since, E is an orthonormal matrix $E'E = I$, where I, is an identity matrix n × n

Therefore,

$$C_Y = \frac{1}{n-1}D \tag{21}$$

The principal component's importance is determined by the variances, with the largest variance corresponding to the first principal component. The eigenvalues and eigenvectors are sorted in descending order, and the orthonormal matrix, E, is created by employing the equivalent eigenvectors in a similar directive. The direct association between the PCs and the actual dataset can be written as:

$$PC_i = \lambda_{1i}x_1 + \lambda_{2i}x_2 + \cdots + \lambda_{ip}x_m \tag{22}$$

where $PC_i$ connotes the $i^{th}$ principal components, x ix the transformed matrix. PCA was used to establish the coefficient of regression weight and it is denoted by $\lambda_{ip}$.

$$PC_i = \lambda_{ip}x_m \tag{23}$$

Take the first eigenvalues the $PC_i$ as the input values and the nodes input layer, respectively. $w_{ij}$ signifies the weights linking the input layers, hidden and $b_{ij}$ signifies the hidden layer biases. The $j^{th}$ component input value in the hidden layer can be stated as:

$$\text{total hidden}_j = \sum_{i=1}^{k} w_{ij}PC_i + b_j \tag{24}$$

Subsequently, the $n^{th}$ component input value in the output layer is written as:

$$total_{output_n} = \sum_{j=1}^{m} \rho_{jn}t_j + \delta_u \tag{25}$$

where m, means the quantity of the nodes of the concealed layers, $t_j$ and $\delta_u$ represent the input values and biases on the hidden layer. $\rho_{jn}$ signifies the weight linking the hidden and the output layers.
The output for hidden and output layers is given as

$$\Lambda_j = f(\text{total hidden}_j = \sum_{i=1}^{k} w_{ij}PC_i + b_j) \tag{26}$$

$$\Omega_n = f(total_{output_n} = \sum_{j=1}^{m} \rho_{jn} t_j + \delta_u ) \tag{27}$$

where, $\Lambda_j$ represents $j^{th}$ node output value in the hidden layer, $f$ is the transfer function utilized for calculation, and $\Omega_n$ is the output value of $n^{th}$ node in the output layer.

Also, the network error is estimated as:

$$\text{error} = \frac{1}{2} \sum_{n=1}^{l} (e_n - \Omega_n)^2 \tag{28}$$

where, $e_n$ represents the desired output value, and $l$ is the nodes number of the output layer.

The back-propagation algorithm is pragmatic as the training algorithm to diminish the network blunder, $\phi$ denotes the learning rate of the network, the revised $\widetilde{w_{ij}}$ and $\widetilde{\rho_{jn}}$ can be expressed as.

$$\widetilde{w_{ij}} = w_{ij} - \phi \frac{\partial error}{\partial w_{ij}} \tag{29}$$

$$\widetilde{\rho_{jn}} = \rho_{jn} - \phi \frac{\partial error}{\partial \rho_{jn}} \tag{30}$$

once the blunder rate or number of training iterations attains the system setting rate, the training becomes stationary and the forecast effect is achieved. Otherwise, the network training will proceed from Eq. (22). Fig. 3a and b depict the modeling process that combines PCA-ANN techniques. The PCA technique is used to minimize the dimensions of QoS variables. The usefulness of the PCA technique and the feasibility of the joint technique (PCA-ANN) is computed by detecting the forecast outcome of the amalgamation of PCA and a single optimal prognosis technique. Figures 3a and b depict the method of analysis used in this study.

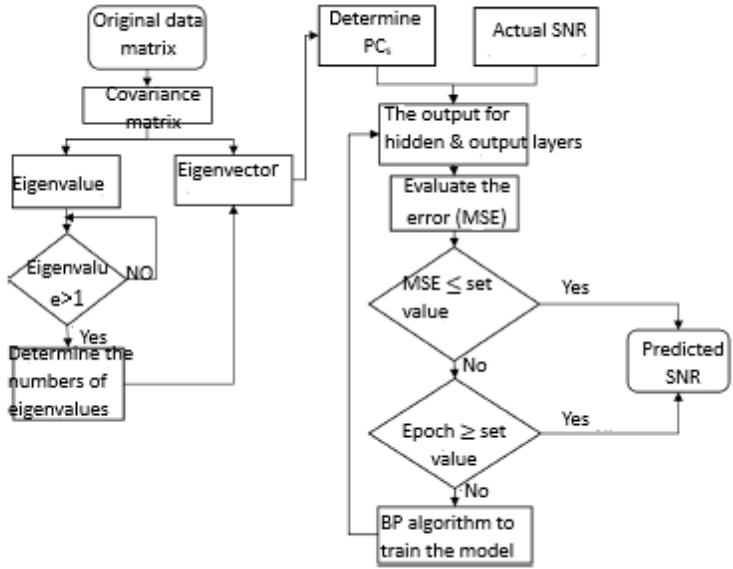

Figure 3 (a): Combining PCA and ANN modeling procedure.

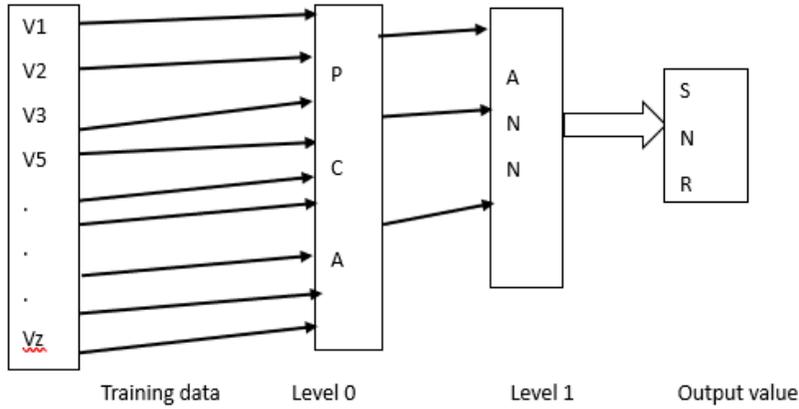

Figure 3b: A two-level hybrid stage

**Table 2:** Evaluation metrics (RMSE, MAPE, MSE, and MAE)

| Meaning | Abbreviation | Explanation | Formula |
|---|---|---|---|
| Mean absolute percentage error | MAPE | Lesser values are improved | $\frac{1}{n}\sum_{i=1}^{n}\left|\frac{Q_a - Q_p}{Q_a}\right|$ |
| Mean absolute error | MAE | Lesser values are improved | $\frac{1}{n}\sum_{i=1}^{n}\left|Q_a - Q_p\right|$ |
| Root mean square error | RMSE | Lesser values are improved | $\sqrt{\frac{\sum_{i=1}^{n}(Q_a - Q_p)^2}{n}}$ |
| Mean square error | MSE | Lesser values are improved | $\frac{\sum_{i=1}^{n}(Q_a - Q_p)^2}{n}$ |

where n signifies the total amount of samples, i connotes the number of terms in the data table. Also, $Q_a$ and $Q_p$ are the measured and predicted values Signal to Noise Ratio.

3.0  Results and Discussion

The operational parameters of FSO systems obtained in Table 1 are selected to advance the link excellence and for examination built based on accessibility and tradition in profitable organizations.

3.1.0 Variation of attenuation with visibility over the study locations

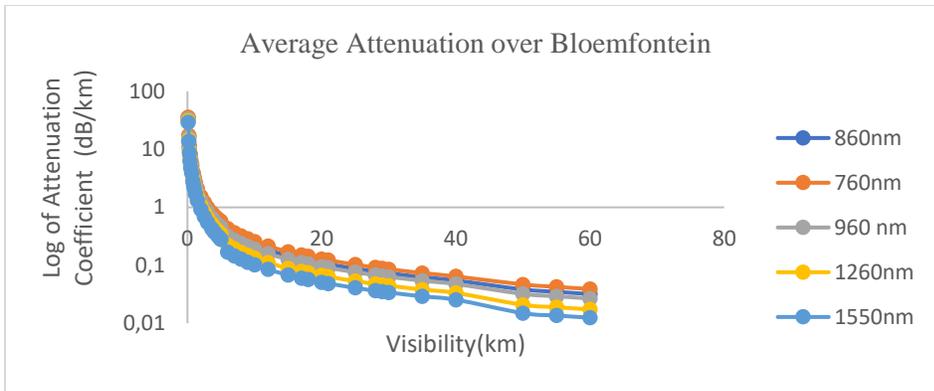

a) Bloemfontein

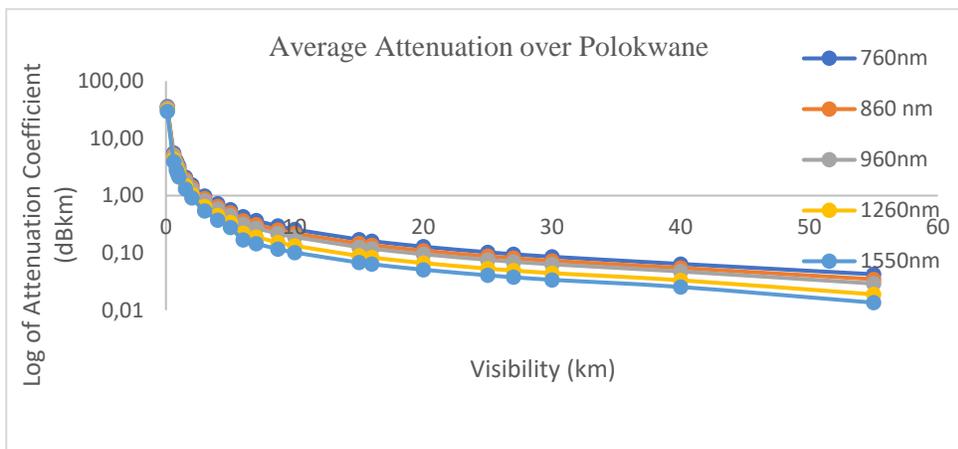

b) Polokwane

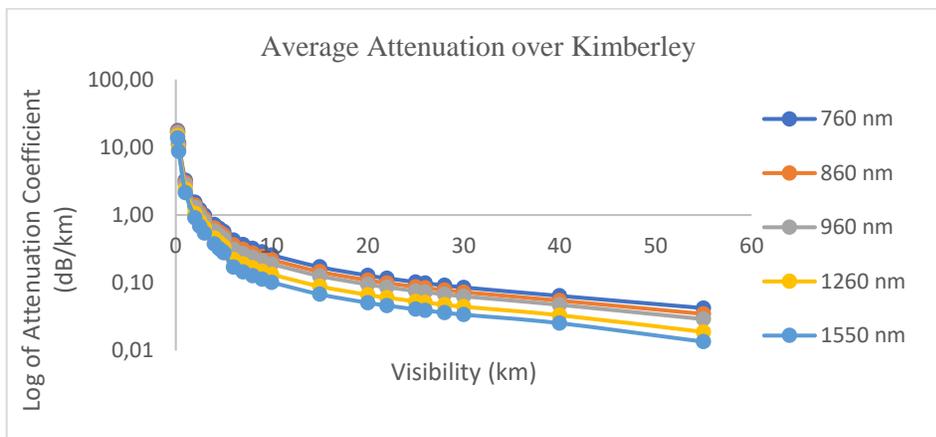

c) Kimberley

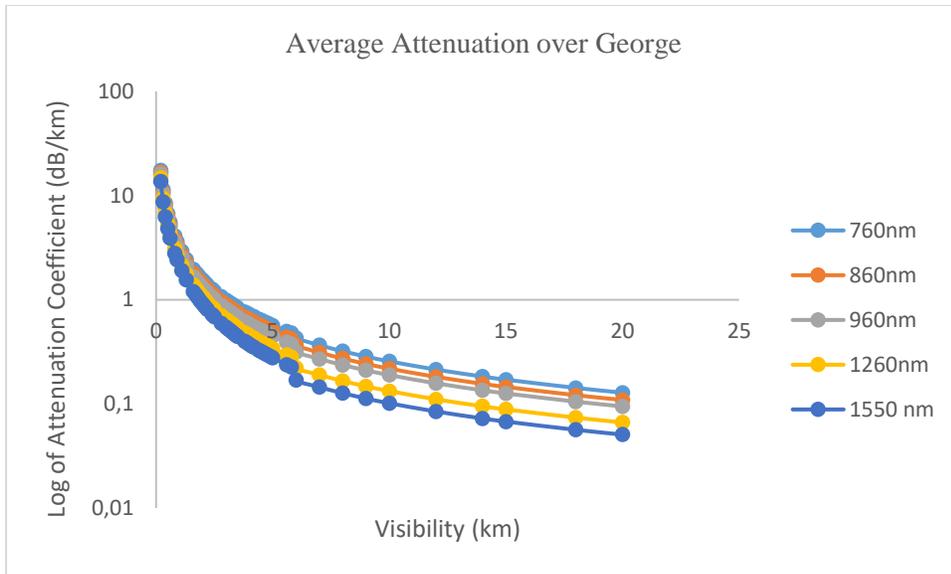

d) George

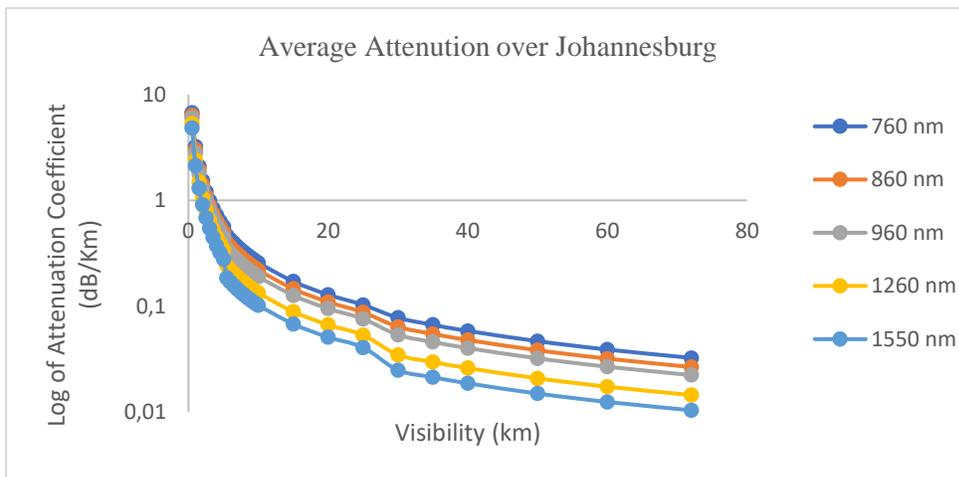

e) Johannesburg

Figures 4. The distributions in variations of attenuation coefficient and visibility at different wavelengths for (a) Bloemfontein (b) Polokwane (c) Kimberley (d) George and (e) Johannesburg.

### 4.1.1 The effect of visibility on fog attenuation coefficients across multiple optical communication wavelengths.

Figures 4a–e displayed the influence of visibility on haze diminution across numerous ocular communication space wavelengths that are regularly applied in free-space optical communication schemes, as attained via the standing Kruse model. It is seen from the figures that the coefficient of attenuation decreases with the increase in visibility and, at the same time, decreases as the wavelength increases across all the locations of the study. This is the result of a reverse association that occurs between the coefficient of attenuation and visibility, as seen in equation (8). As an example, at a wavelength of 1550 nm and visibility of 1 km, the values of the coefficient of attenuation across all the locations, Polokwane, Kimberley, Johannesburg, Bloemfontein, and George, are 2.13, 2.13, 0.3801, 2.13,

and 2.13 dB/km, correspondingly, while at the same visibility with a wavelength of 760 nm, the values of the coefficient of attenuation across the same locations are 3.240, 3.010, 0.543, 3.236, and 2.923 dB in that order. The percentage of declination while transmitting an FSO signal from 760 nm to 1550 nm at 1km across the locations of Polokwane, Kimberley, Johannesburg, Bloemfontein, and George is 34.26%, 29.24%, 29.91%, 34.18%, and 27.12%, respectively. From the result obtained, it could be inferred that less of an FSO signal is attenuated while transmitting at a higher wavelength. Thus, telecommunications service providers are recommended to use the higher wavelengths for optimal linkage efficacy and QoS.

**4.1.2** Variation of signal-to-noise ratio with attenuation over the study locations

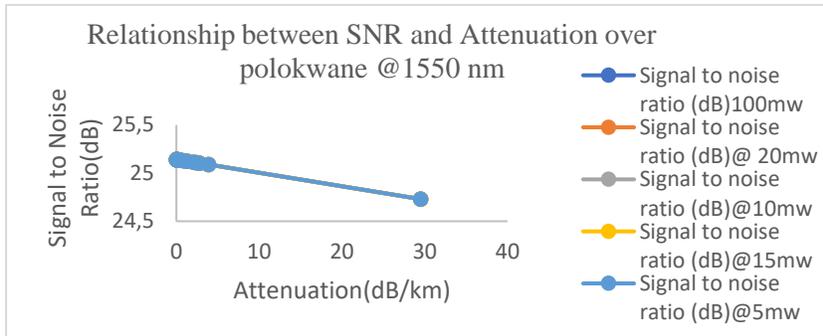

(a) Polokwane

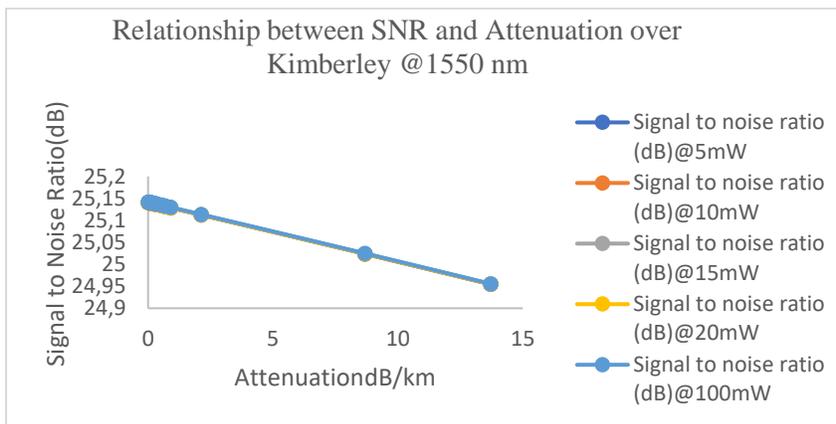

(b) Kimberley

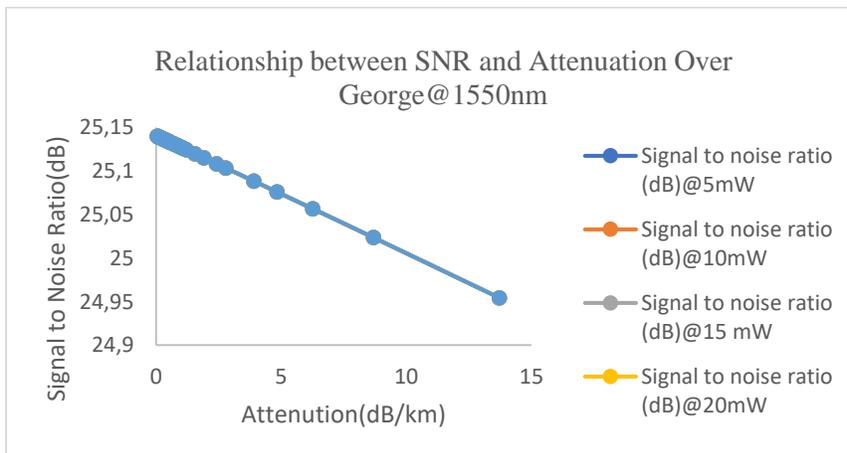

(C) George

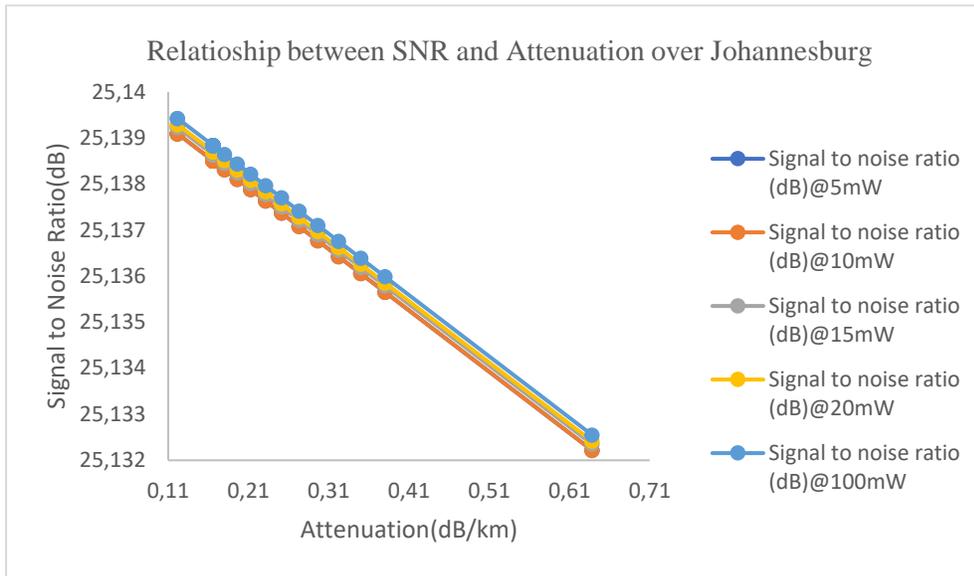

(d) Johannesburg

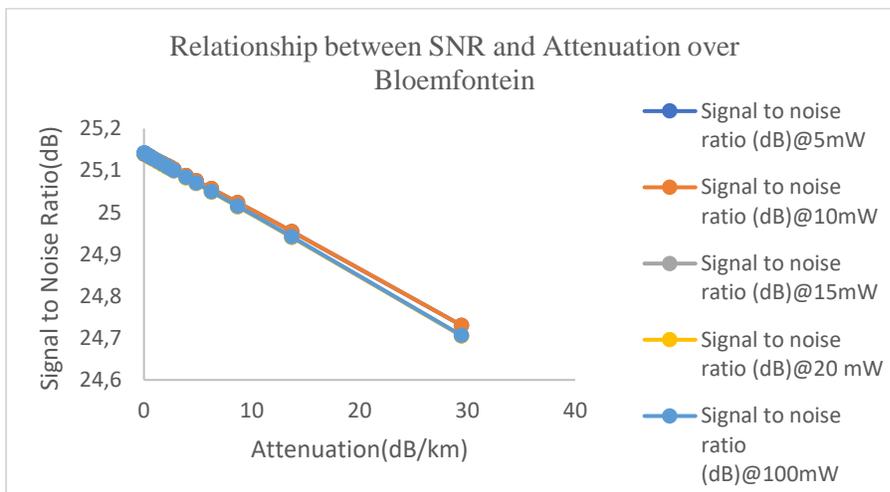

(e) Bloemfontein

Figures.5a-e. Variations of Attenuation Coefficient with Signal-to-noise ratio at power rating for (a) Polokwane (b)Kimberley (c) George (d) Johannesburg (e) Bloemfontein.

4.1.2 Relationship between SNR and Attenuation Coefficient

In determining the worth of an FSO link, the SNR is an imperative factor of the FSO linkage, and it has a straight adverse association with the coefficient of diminution, as seen in equation (11). Figures 5(a-e) show that SNR rises with the reduction in diminution; the reduction in SNR is very sharp. The upsurge of power conveyed leads to an upsurge in SNR for the reason that SNR is a ratio of signal power and noise power. Likewise, at 100 mW of transmitted power with an attenuation coefficient of 4.10 dB/km, the SNR values over the locations Polokwane, Bloemfontein, George, and Kimberley are found to be 25.09, 25.08, 25.09, and 25.09 dB, respectively. of Besides, at 5 mW of conveyed power with a coefficient of diminution 5.55 dB/km, the SNR values in the locations Polokwane, Bloemfontein, George, and Kimberley are found to be 25.070, 25.080, 25.071, and 25.054 dB, respectively. Also, in Johannesburg, at 0.61 dB/km of attenuation coefficient, the SNR is 25.133 dB, which validates the concept that the SNR rises as the conveyed power rises. As the coefficient of diminution rises from 4.10 dB/km to 5.55 dB/km, there is an average reduction of 0.08% in the values of SNR found across all the locations. The implication of this is that at a small value of the diminution coefficient and at a high signal-to-noise ratio value, a good quality of service could be achieved.

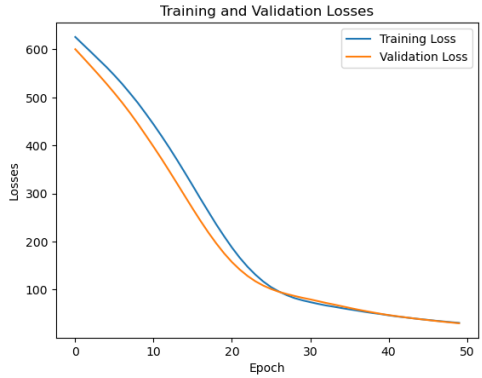

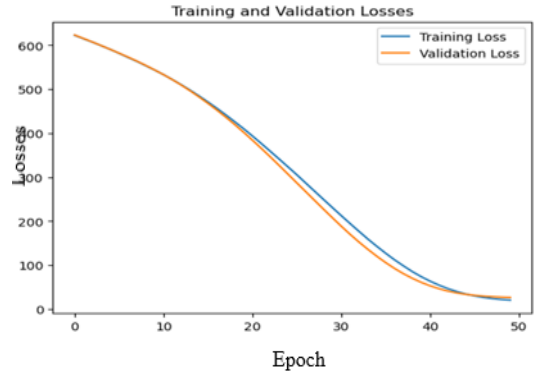

(a) Johannesburg                                  (b) Polokwane

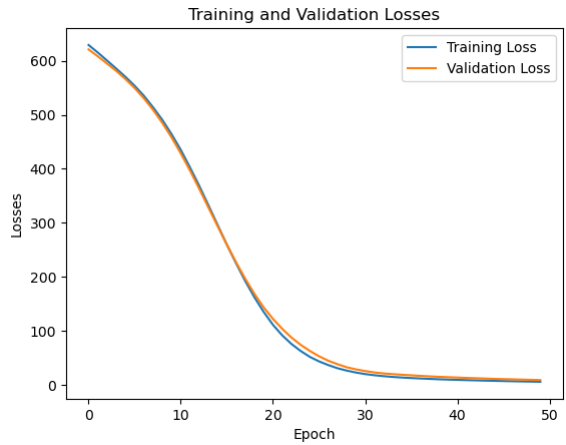

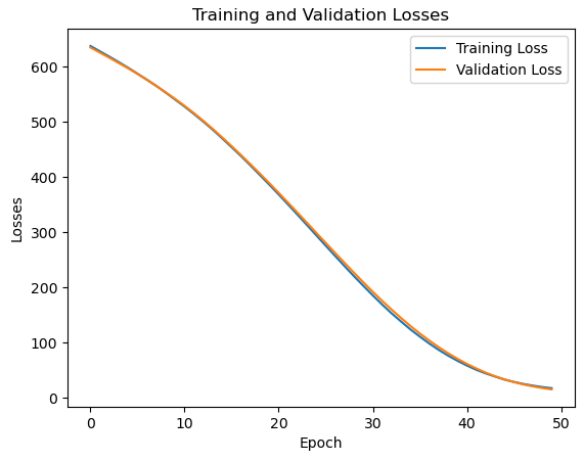

(c) George                                        (d) Kimberley

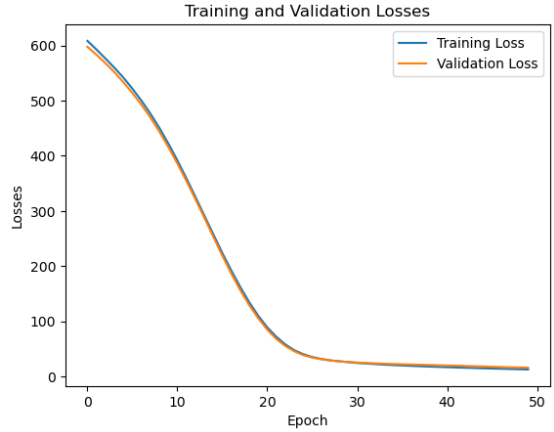

(e) Bloemfontein

Figures 6a-e. Analysis of the training and validation losses for (a) Johannesburg (b) Polokwane (c) George (d) Kimberley (e) Bloemfontein

4.2.0 Validation and Training Loss

Figures 6(a-e) show the contrast between the training and the validation loss across all the locations of the study: Johannesburg, Polokwane, George, Kimberley, and Bloemfontein. In Figure 6a, it is observed that at epoch 10, the training and the validation loss is 466.48 and 421.25 dB, respectively, while at epoch 50, the training and the validation loss is 30.70 and 29.99 dB. Moreover, in Figure 5b, the training and validation losses at epoch 10 is 491.51 and 448.98 dB, respectively. At 50 epochs, the training and validation losses is 34.97 and 33.99 dB, respectively. The same trend could be observed in all other locations under investigation (Figures 6a-e) in this study but with different validation losses. Therefore, it is rational to deduce that more epochs will result in lesser losses and higher QoS prediction accuracy based on the research method. Nevertheless, the quantity of epochs rises as the training phase rises. Arbitrating from Figures 6(a-e), there is no significant difference between the training and the validation losses; therefore, the proposed model is a good fit for predicting the QoS.

**Table3. Analysis of evaluation metrics**

| PERFORMANCE CRITERION METRICS | | | | |
|---|---|---|---|---|
| Locations | MAPE | MAE | RMSE | MSE |
| Bloemfontein | 0.1437 | 3.6083 | 4.1611 | 17.3140 |
| Polokwane | 0.1977 | 4.9641 | 5.6380 | 31.7860 |
| Johannesburg | 0.2131 | 5.3567 | 6.1198 | 37.4500 |
| George | 0.1101 | 2.7643 | 3.4714 | 12.0507 |
| Kimberley | 0.2329 | 5.8517 | 6.6139 | 43.7460 |

Rendering to the estimation outcomes of the proposed model, the assessment presentation reflected by MAPE, MAE, MSE, and RMSE in all the locations, Johannesburg, Polokwane, George, Kimberley, and Bloemfontein is seen in Table 3. The MAPE values across the locations, Johannesburg, Polokwane, George, Kimberley, and Bloemfontein, were 0.2131, 0.1977, 0.1101, 0.2329, and 0.1437, respectively. It is perceived from Table 3 that when the MAPE value is smaller, it is accompanied by a smaller MAE value, a smaller RMSE value, and a larger R-value, which signifies a far improved forecast consequence. The proposed model performs better in predicting the QoS. This is because the lesser the values of MAPE, the better the performance of the proposed model.

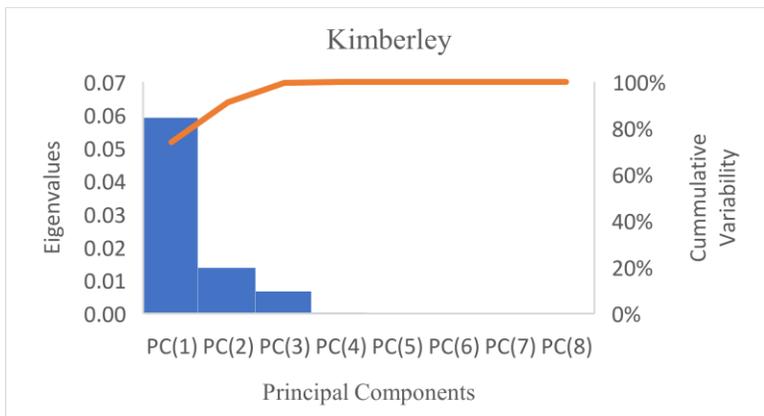

a) Kimberley

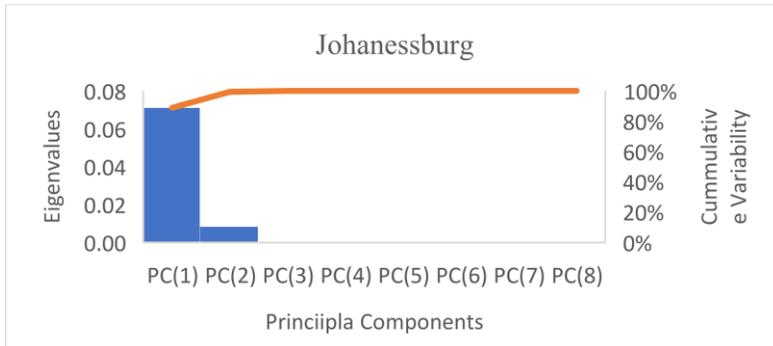

b) Johannesbourg

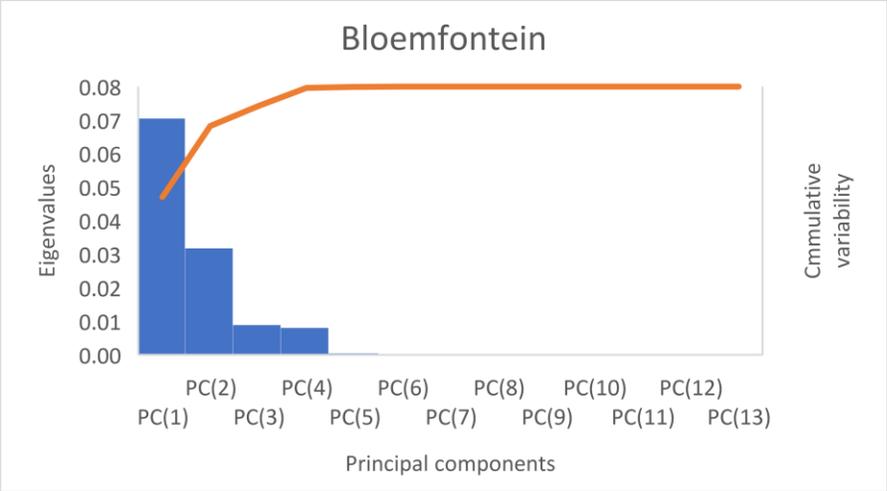

c) Bloemfontein

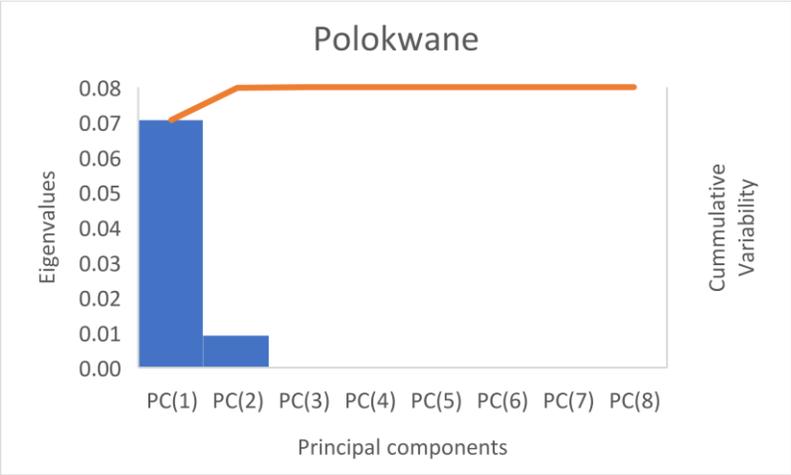

d) Polokwane

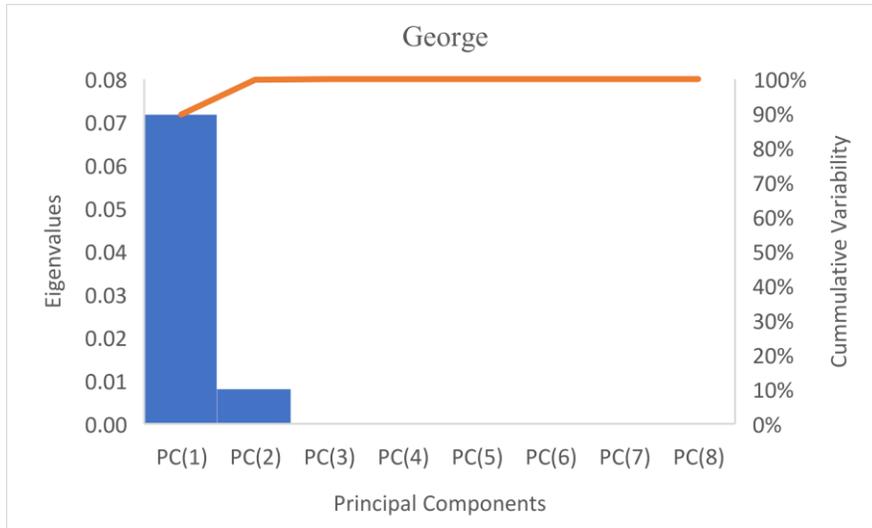

e) George

Figures 7a-e. Scree plot of eigenvalues of different PCs for (a) Kimberley (b) Johannesburg (c) Bloemfontein (d) Polokwane (e) George.

Figure 7(a–e) shows the screeplot of the eigenvalues, principal components, and cumulative variability of the different locations under study. The plot is necessary to determine the number of factors to be retained under the number of PCs to keep in the PCA. Usually, the eigenvalue 1 indicates that the principal component (PC) comprehends the entirety of the unique variable data that has a solid clarification of the actual QoS variable and could be reserved. It is seen in all the locations that the eigenvalues of PC (1) and PC (2) are greater than 1; therefore, they are reserved as the new QoS variable. The PC (1) has the largest cumulative variance values across all the locations: Kimberley, Polokwane, Johannesburg, Bloemfontein, and George, and they are found to be 74.0%, 88.4%, 88.9%, 88.8%, and 89.7%, respectively.

## 4.0 Conclusion

This study has presented a hybrid PCA-ANN model for predicting the QoS in FSO communication systems using a back propagation neural network and also the influence of atmospheric attenuation coefficient on the sustainability of signal QoS. The results show that the atmospheric attenuation coefficient reduces as the optical link distance increases while the atmospheric attenuation coefficient decreases as the optical wavelength increases. Moreover, the signal-to-noise ratio increases with a decrease in the atmospheric attenuation coefficient. The current hybrid artificial neural network (ANN) model outperformed the earlier empirical formula or artificial neural network (ANN) models, as evidenced by a significantly stronger correlation coefficient when compared to the observed stability values. Based on this investigation, it was shown that more effective QoS may be efficiently produced to satisfy Customers' Service Level Agreements (SLAs) by integrating PCA with ANN for signal modeling. This will enable the achievement of a single valuable product, which will result in a satisfying performance. Ultimately, the combined outcome will give FSO systems engineers the technologies needed to attain the link margin for the best quality of service during extreme subtropical weather periods.

Declaration of Competing Interest

The authors state that none of the work presented in this study may have been influenced by any known conflicting financial interests or personal ties.

Acknowledgment

The Authors would also like to acknowledge South African Weather Services (SAWS) for making the meteorological data available for the research.